\documentclass[
aps,%
12pt,%
final,%
notitlepage,%
oneside,%
onecolumn,%
nobibnotes,%
nofootinbib,%
superscriptaddress,%
noshowpacs,%
centertags]%
{revtex4}

\begin{document}
\selectlanguage{english}

\title{The Spectral Type of the Ionizing Stars and the Infrared Fluxes of HII Regions}
\author{A.~P.~Topchieva}
\email{ATopchieva@inasan.ru}
\affiliation{Institute of Astronomy, Russian Academy of Sciences, str. Pyatnitskaya 48, Moscow, 119017 Russia}

\author{M.~S.~Kirsanova}
\affiliation{Institute of Astronomy, Russian Academy of Sciences, str. Pyatnitskaya 48, Moscow, 119017 Russia}
\affiliation{Barkhatova Kourovka Astronomical Observatory, Yeltsin Ural Federal University,pr. Lenina 51, Yekaterinburg, 620000 Russia}

\author{A.~M.~Sobolev}
\affiliation{Barkhatova Kourovka Astronomical Observatory, Yeltsin Ural Federal University,pr. Lenina 51, Yekaterinburg, 620000 Russia}


\begin{abstract}
The 20~cm radio continuum fluxes of 91 HII regions in a previously compiled catalog have been determined. The spectral types of the ionizing stars in 42 regions with known distances are estimated. These spectral types range from B0.5 to O7, corresponding to effective temperatures of 29 000-37 000 K. The dependences of the infrared (IR) fluxes at 8, 24, and 160 $\mu$m on the 20~cm flux are considered. The IR fluxes are used as a diagnostic of heating of the matter, and the radio fluxes as measurements of the number of ionizing photons. It is established that the IR fluxes grow approximately linearly with the radio flux. This growth of the IR fluxes probably indicates a growth of the mass of heated material in the envelope surrounding the HII region with increasing effective temperature of the star.
\end{abstract}

\maketitle

\section{Introduction}\label{sec:intro}

Massive stars are fairly rare in the Galaxy, but they exert a strong influence on the state of the interstellar medium (ISM). As a rule, the ages of massive stars do not exceed 10 million years --- much less than the age of the Milky Way --- so that their locations are good tracers of regions of active star formation and the structure of the Galaxy (e.g.,~\cite{Russeil, Reid, Anderson_WISE}). One of the main factors in the influence of massive stars on the ISM is their bright ultraviolet (UV) radiation, leading to the formation of surrounding regions of ionized hydrogen (HII regions)~\cite{Shu} with expanding ionization and shock fronts. There is a transition region between such an HII region and the surrounding gas of the host molecular cloud, forming a molecular photodissociation region (PDR). The hydrogen in this region is predominantly neutral, but the degree of ionization of the gas is about 0.01\% due to the ionization of carbon. Since the density of the heated gas and dust in a PDR is usually high (see, e.g., the review D. J. Hollenbach и A. G. G. M. Tielens~\cite{Hollenbach_99}), PDRs are luminous in the infrared (IR). Results from the ``Spitzer'' IR telescope reveal the presence of thousands of nebulae in the Galactic disk whose shapes are ring-like at 8 $\mu$m~\cite{2006ApJ...649..759C, 2007ApJ...670..428C, Topchieva}. Deharveng et al.~\cite{Deharveng_10} showed that most of these ring nebulae are IR reflections of HII regions. Polycyclic aromatic hydrocarbons (PAHs) radiate at 8 $\mu$m in PDRs heated by the UV radiation of massive stars. Longer-wavelength IR emission in HII regions is radiated by larger heated dust grains~\cite{Draine, 2007ApJ...657..810D}. It is interesting that 24 $\mu$m emission (the photometric range of the ``Spitzer'' telescope) is observed inside the 8 $\mu$m rings, while longer-wavelength emission (70, 160, 250, and sometimes 350 $\mu$m) is coincident with the ring observed at 8 $\mu$m~\cite{Topchieva}. Studies of the spatial distribution of the IR emission in HII regions can provide information about the dust dynamics and the destruction of dust by UV radiation from massive stars (e.g., E. Kruegel~\cite{Kruegel}, J. S. Mathis~\cite{Mathis}, Pavlychenkov et al.~\cite{2013ARep...57..573P}, Akimkin et al.~\cite{2015MNRAS.449..440A,Akimkin_17}, Murga et al.~\cite{Murga_16}).

The spectral type and spectral energy distributions of the ionizing stars are key parameters in any model for the dynamics and destruction of dust and PAHs in the ISM~\cite{Draine, Murga_16}. Unfortunately, such data are available for only a relatively small fraction of the IR nebulae observed with the ``Spitzer'' telescope, which were studied before the launch of Spitzer. However, the spectral types of young, massive stars in HII regions can be estimated based on archival data on the radio emission of the heated ionized gas (e.g.,~\cite{Dirienzo}). This makes it possible to obtain the parameters of the ionizing sources in large numbers of HII regions.

During the expansion of an HII region, there is an increase in the mass of the surrounding neutral, dense envelope due to the action of shocks~\cite{Hosok}. However, no analytical and numerical studies of time variations of the IR fluxes in HII regions have been conducted. In our current study, we have collected and analyzed observational data on the radio and IR fluxes, which can be used as a basis for numerical simulations, for example, using the MARION model~\cite{Kirsanova_09,2015MNRAS.449..440A}. We plan to compare these data with results from MARION computations (in a one-dimensional geometry) in the future. Topchieva et al.~\cite{Topchieva} earlier compiled a catalog of HII regions whose shapes are close to circular in the IR. The analysis of available IR data for objects in this catalog is described in~\cite{2017ASPC..510...98T, Topchieva1}. All the analysis in our present study is based on~\cite{2017ASPC..510...98T, Topchieva1} and the WISE Catalog of Galactic HII Regions V2.0\footnote{astro.phys.wvu.edu/wise/}~\cite{Anderson_WISE}.

\section{Radio continuum flux at 20~cm}\label{sec:data}

\begin{figure}[h!]
\includegraphics[width=0.35\linewidth]{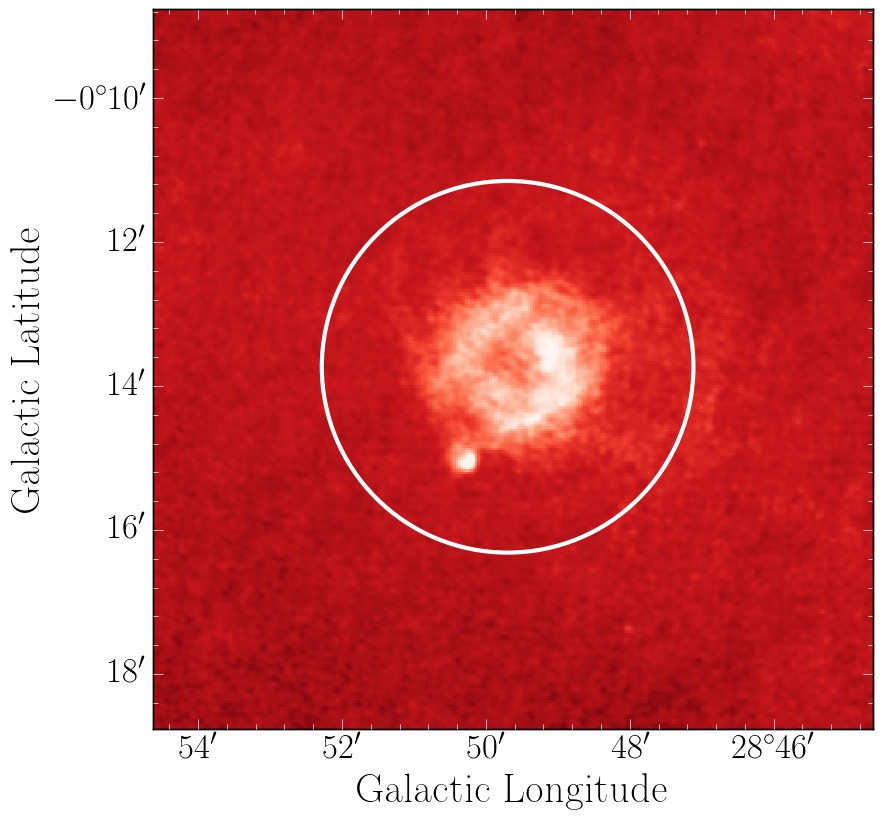}
\includegraphics[width=0.35\linewidth]{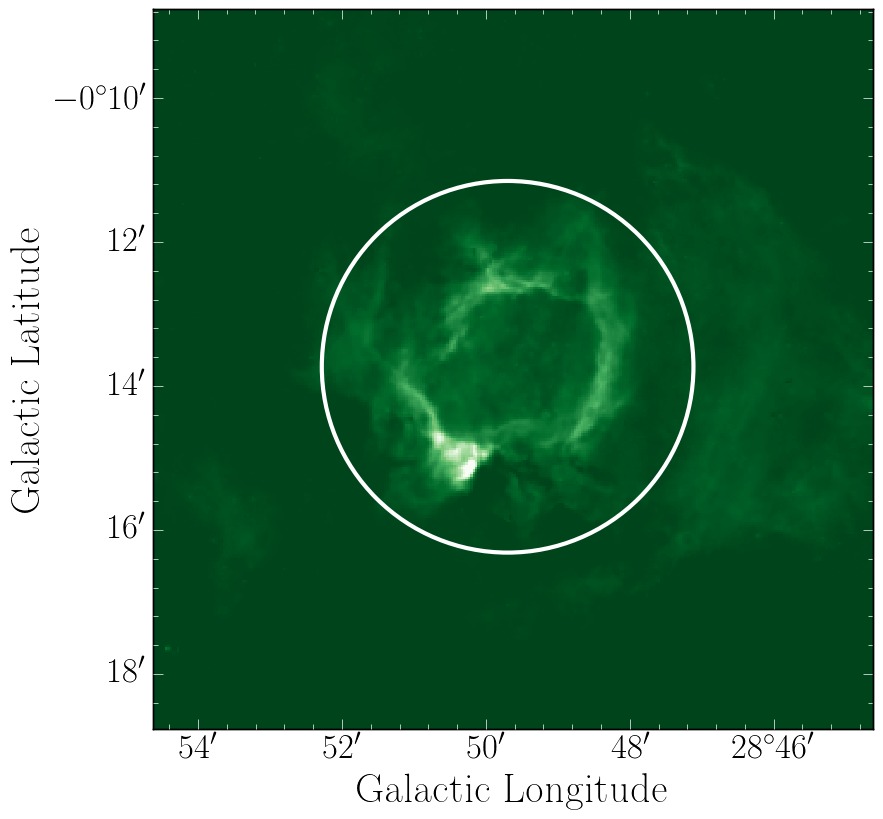}
\caption{The IR nebula N49. Shown are continuum images at 20~cm (left) and 8 $\mu$m (right). The white circle shows the aperture within which we determined the 20~cm flux.}
\label{fig:q}
\end{figure}

The catalog~\cite{Topchieva} includes 99 IR nebulae associated with HII regions. We calculated the 20~cm radio fluxes ($F_{20cm}$) of 91 of these HII regions for which data were presented in the New~GPS~20~cm\footnote{http://third.ucllnl.org/gps/} and by Condon et al.~\cite{Condon}. We calculated the fluxes within an aperture determined from the 8 $\mu$m data in~\cite{Topchieva, Topchieva1}, which was used for all the wavelengths. We determined the radio fluxes of these 91 objects from the catalog~\cite{Topchieva} in this way. Figure~\ref{fig:q} shows the utilized aperture for the nebula N49 as an example (notation is from the catalog~\cite{2006ApJ...649..759C}). This 20~cm image was taken from the MAGPIS radio-image database~\cite{Helfand_MAGPIS}.

Determining the spectral type of the ionizing stars based on the radio continuum flux requires knowledge of the distances to these stars. We did not determine these distances ourselves, instead using data from catalogs of HII regions, primarily~\cite{Anderson_WISE}, which presents more than 8000 objects based on IR data. We also used the distance estimates of Churchwell et al.~\cite{2006ApJ...649..759C, 2007ApJ...670..428C}.

\newpage
\begin{longtable}{|l|l|r|r|r|}
\caption{20~cm radio continuum fluxes for objects from the catalog of Topchieva et al.~\cite{Topchieva}. Objects in the catalog were taken $^{1}$R. H. Becker et. al.~\cite{1994ApJS...91..347B}, $^{2}$R.~J. Simpson et. al.~\cite{2012MNRAS.424.2442S}, $^{3}$E.~Churchwell et. al.~\cite{2006ApJ...649..759C}, $^{4}$J. S. Urquhart et. al.~\cite{2009A&A...507..795U}, $^{5}$M. P. Egan et. al.~\cite{2003yCat.5114....0E}.}
\label{tab:catal8mkm}\\ \hline
$N\textsuperscript{\underline{o}}$& Name & $l_{gal}$, $^{\circ}$ & $b_{gal}$, $^{\circ}$ & $F_{20}$, Ян\\ \hline
\endfirsthead \hline
\multicolumn{5}{|c|}{\scriptsize\slshape(to be continued)} \\ \hline
$N\textsuperscript{\underline{o}}$& Name & $l_{gal}$, $^{\circ}$ & $b_{gal}$, $^{\circ}$ & $F_{20}$, Ян\\ \hline
\endhead \hline
\multicolumn{5}{|c|}{\scriptsize\slshape to be continued} \\ \hline
\endfoot \hline
\endlastfoot

1  &  CN67$^{3}$                   & 5.526      & 0.037        &0.44$\pm$0.14      \\
2  &  CN77$^{3}$                   & 6.139      & --0.640      &1.91$\pm$0.19      \\
3  &  CN79$^{3}$                   & 6.202      & --0.334      &2.01$\pm$0.23      \\
4  &  CN111$^{3}$                  & 8.311      & --0.086      &2.48$\pm$0.33      \\
5  &  MWP1G008430--002800S$^{2}$   & 8.431      & --0.276      &0.22$\pm$0.03      \\
6  &  CN116$^{3}$                  & 8.476      & --0.277      &0.14$\pm$0.02      \\
7  &  N4$^{3}$                     & 11.893     & 0.747        &4.74$\pm$0.49      \\
8  &  MWP1G012590--000900S$^{2}$   & 12.595     & --0.090      &0.23$\pm$0.03      \\
9  &  MWP1G012630--000100S$^{2}$   & 12.633     & --0.017      &0.19$\pm$0.04      \\
10 &  N8$^{3}$                     & 12.805     & --0.312      &0.11$\pm$0.01      \\
11 &  MWP1G013213--001410$^{2}$    & 13.213     & --0.141      &1.68$\pm$0.10      \\
12 &  N13$^{3}$                    & 13.899     & --0.014      &0.59$\pm$0.07      \\
13 &  N14$^{3}$                    & 14.000     & --0.136      &7.19$\pm$0.43      \\
14 &  G014.175+0.024$^{6,1}$       & 14.175     & 0.022        &0.25$\pm$0.02      \\
15 &  MWP1G014210--001100S$^{2}$   & 14.206     & --0.110      &0.25$\pm$0.01      \\
16 &  MWP1G014390--000200S$^{2}$   & 14.388     & --0.024      &0.48$\pm$0.04      \\
17 &  MWP1G014480--000000S$^{2}$   & 14.490     & 0.022        &0.70$\pm$0.03      \\
18 &  MWP1G016390--001400S$^{2}$   & 16.391     & --0.138      &0.16$\pm$0.03      \\
19 &  MWP1G016429--001984$^{2}$    & 16.431     & --0.201      &2.30$\pm$0.28      \\
20 &  MWP1G016560+000056$^{2}$     & 16.560     & 0.002        &0.12$\pm$0.03      \\
21 &  MWP1G017626+000493$^{2}$     & 17.625     & 0.048        &0.09$\pm$0.05      \\
22 &  TWKK1$^{4}$                  & 17.805     & 0.074        &0.03$\pm$0.01      \\
23 &  N20$^{3}$                    & 17.918     & --0.687      &0.31$\pm$0.09      \\
24 &  MWP1G018440+000100S$^{2}$    & 18.442     & 0.013        &0.09$\pm$0.01      \\
25 &  MWP1G018580+003400S$^{2}$    & 18.582     & 0.345        &0.31$\pm$0.03      \\
26 &  N23$^{3}$                    & 18.679     & --0.237      &0.79$\pm$0.07      \\
27 &  MWP1G018743+002521$^{2}$     & 18.748     & 0.256        &0.37$\pm$0.03      \\
28 &  MWP1G020387--000156$^{2}$    & 20.388     & --0.017      &0.16$\pm$0.05      \\
29 &  MWP1G02100--000500S$^{2}$    & 21.005     & --0.054      &0.17$\pm$0.01      \\
30 &  N28$^{3}$                    & 21.351     & --0.137      &0.12$\pm$0.09      \\
31 &  N31$^{3}$                    & 23.842     & 0.098        &0.44$\pm$0.04      \\
32 &  MWP1G023849--001251$^{2}$    & 23.848     & --0.127      &0.14$\pm$0.03      \\
33 &  MWP1G023881--003497$^{2}$    & 23.881     & --0.350      &0.02$\pm$0.04      \\
34 &  N32$^{3}$                    & 23.904     & 0.070        &0.58$\pm$0.07      \\
35 &  MWP1G023982--001096$^{2}$    & 23.982     & --0.110      &0.07$\pm$0.02      \\
36 &  MWP1G024019+001902$^{2}$     & 24.043     & 0.204        &0.02$\pm$0.01      \\
37 &  MWP1G024149--000060$^{2}$    & 24.153     & --0.011      &0.07$\pm$0.02      \\
38 &  N33$^{3}$                    & 24.215     & --0.044      &0.82$\pm$0.02      \\
39 &  TWKK3$^{4}$                  & 24.424     & 0.220        &0.73$\pm$0.02      \\
40 &  TWKK2$^{4}$                  & 24.460     & 0.506        &0.03$\pm$0.01      \\
41 &  MWP1G024500--002400$^{2}$    & 24.502     & --0.237      &0.13$\pm$0.01      \\
42 &  MWP1G024558--001329$^{2}$    & 24.558     & --0.133      &0.34$\pm$0.02      \\
43 &  MWP1G024649--001131$^{2}$    & 24.651     & --0.078      &0.07$\pm$0.01      \\
44 &  MWP1G01024699--001486$^{2}$  & 24.700     & --0.148      &0.20$\pm$0.03      \\
45 &  MWP1G024731+001580$^{2}$     & 24.736     & 0.158        &0.19$\pm$0.01      \\
46 &  MWP1G024920+000800$^{2}$     & 24.922     & 0.078        &0.32$\pm$0.02      \\
47 &  MWP1G025155+000609$^{2}$     & 25.155     & 0.061        &0.50$\pm$0.06      \\
48 &  MWP1G025723+00058$^{2}$      & 25.724     & 0.058        &0.25$\pm$0.02      \\
49 &  MWP1G025730--000200S$^{2}$   & 25.726     & --0.027      &0.06$\pm$0.01      \\
50 &  N42$^{3}$                    & 26.329     & --0.071      &0.38$\pm$0.09      \\
51 &  N43$^{3}$                    & 26.595     & 0.095        &0.55$\pm$0.11      \\
52 &  MWP1G026720+001700S$^{2}$    & 26.722     & 0.173        &0.29$\pm$0.01      \\
53 &  G027.492+0.192$^{6}$         & 27.496     & 0.197        &2.21$\pm$0.05      \\
54 &  MWP1G02671+00300S$^{2}$      & 27.613     & 0.028        &0.06$\pm$0.01    \\
55 &  MWP1G027905--000079$^{2}$    & 27.904     & --0.009      &0.12$\pm$0.02    \\
56 &  G027.9334+00.2056$^{6,5}$    & 27.931     & 0.205        &0.10$\pm$0.02    \\
57 &  MWP1G027981+000753$^{2}$     & 27.981     & 0.073        &0.42$\pm$0.04    \\
58 &  MWP1G028160--000300S$^{2}$   & 28.160     & --0.046      &0.09$\pm$0.02    \\
59 &  N49$^{3}$                    & 28.827     & --0.229      &2.46$\pm$0.27    \\
60 &  MWP1G029136--001438$^{2}$    & 29.134     & --0.144      &0.09$\pm$0.02    \\
61 &  N51$^{3}$                    & 29.156     & --0.259      &1.04$\pm$0.45    \\
62 &  MWP1G030020--000400S$^{2}$   & 30.022     & --0.041      &0.67$\pm$0.05    \\
63 &  MWP1G030250+002413$^{2}$     & 30.251     & 0.240        &0.17$\pm$0.05    \\
64 &  MWP1G03080+001100S$^{2}$     & 30.378     & 0.111        &0.21$\pm$0.04    \\
65 &  MWP1G030381--001074$^{2}$    & 30.381     & --0.109      &0.42$\pm$0.08    \\
66 &  MWP1G031066+000485$^{2}$     & 31.071     & 0.049        &0.59$\pm$0.01    \\
67 &  MWP1G032057+000783$^{2}$     & 32.055     & 0.076        &0.35$\pm$0.10    \\
68 &  N55$^{3}$                    & 32.101     & 0.091        &0.43$\pm$0.16    \\
69 &  MWP1G032731+002120$^{2}$     & 32.730     & 0.212        &0.07$\pm$0.07    \\
70 &  N57$^{3}$                    & 32.761     & --0.149      &0.18$\pm$0.04    \\
71 &  N60$^{3}$                    & 33.815     & --0.149      &0.21$\pm$0.06    \\
72 &  MWP1G034088+004405$^{2}$     & 34.087     & 0.441        &0.36$\pm$0.11    \\
73 &  MWP1G034680+000600S$^{2}$    & 34.684     & 0.067        &0.14$\pm$0.01    \\
74 &  N67$^{3}$                    & 35.544     & 0.012        &0.42$\pm$0.08    \\
75 &  MWP1G037196--004296$^{2}$    & 37.195     & --0.429      &0.08$\pm$0.06    \\
76 &  MWP1G037261--000809$^{2}$    & 37.258     & --0.078      &0.34$\pm$0.01    \\
77 &  MWP1G037349+006876$^{2}$     & 37.351     & 0.688        &0.32$\pm$0.07    \\
78 &  N70$^{3}$                    & 37.750     & --0.113      &0.60$\pm$0.01    \\
79 &  G038.550+1648$^{6}$          & 38.551     & 0.162        &0.23$\pm$0.01    \\
80 &  N73$^{3}$                    & 38.736     & --0.140      &0.24$\pm$0.14    \\
81 &  N78$^{3}$                    & 41.228     & 0.169        &0.02$\pm$0.01    \\
82 &  G041.378+0.035$^{6,1}$       & 41.378     & 0.034        &0.10$\pm$0.06    \\
83 &  N79$^{3}$                    & 41.513     & 0.031        &1.03$\pm$0.05    \\
84 &  TWKK4$^{4}$                  & 41.595     & 0.160        &0.01$\pm$0.01    \\
85 &  N80$^{3}$                    & 41.932     & 0.033        &0.57$\pm$0.22    \\
86 &  N89$^{3}$                    & 43.739     & 0.114        &0.15$\pm$0.06    \\
87 &  N90$^{3}$                    & 43.774     & 0.060        &0.64$\pm$0.27    \\
88 &  MWP1G045540+000000S$^{2}$    & 45.544     & --0.005      &0.17$\pm$0.01    \\
89 &  N96$^{3}$                    & 46.949     & 0.371        &0.20$\pm$0.04    \\
90 &  N98$^{3}$                    & 47.027     & 0.218        &0.43$\pm$0.37    \\
91 &  MWP1G048422+001173$^{2}$     & 48.422     & 0.116        &0.28$\pm$0.10    \\
\end{longtable}

\newpage
\scriptsize
\begin{longtable}{|l|r|r|r|r|r|r|r|r|r|r|}
\caption{Distances, sizes, and electron densities of the objects. The ordinal number of each object corresponds to the number from Table~\ref{tab:catal8mkm}. $S$ is the diameter. The distances $D$ were taken from~\cite{Anderson_WISE}. The spectral types in columns (6-9) correspond to~\cite{Smith}, and those in columns 10 and 11 to~\cite{Vacca}.}
\label{tab:mainres} \\
\hline
$N\textsuperscript{\underline{o}}$& S,  & D, & n$_{\rm e}$,  & $logQ_{\rm Ly}$, $сек^{-1}$ & T$_{\rm eff}^{\rm V}$,& Sp.typ.$^{\rm V}$ & T$_{\rm eff}^{\rm III}$, & Sp.typ.$^{\rm III}$ & T$_{\rm eff}^{\rm V*}$, & Sp.typ.$^{\rm V*}$\\
    & pc & kpc&cm$^{-3}$ &($logQ_{\rm min}$, $logQ_{\rm max}$)& K  &  & K  & & K  &\\ \hline
\endfirsthead \hline
\multicolumn{11}{|c|}{\scriptsize\slshape(to be continued)} \\ \hline
$N\textsuperscript{\underline{o}}$& S,  & D,  & n$_{e}$,  & $logQ_{\rm Ly}$, $s^{-1}$ & T$_{eff}^{\rm V}$, K & Sp.typ.$^{\rm V}$ & T$_{eff}^{\rm III}$, K & Sp.typ.$^{\rm III}$ & T$_{eff}^{\rm V*}$, K & Sp.typ.$^{\rm V*}$\\
    & pc&kpc&cn$^{-3}$    &($logQ_{\rm min}$, $logQ_{\rm max}$)& & & & & &\\ \hline
\endhead \hline
\multicolumn{11}{|c|}{\scriptsize\slshape to be continued} \\ \hline
\endfoot \hline
\endlastfoot
 2& 2.8$\pm$2.8        &4.3$\pm$1.0   &255.98     &48.44(47.36,48.72) &35049.30   &O9/O9.5  &30999.68   &B0/B0.5  &34371.42   &O8/O9      \\    
 4& 5.1$\pm$2.5        &5.3$\pm$0.5   &127.23     &48.73(47.66,49.01) &37287.32   &O8.5/O9  &32656.72   &O9.5/B0  &37686.93   &O7/O7.5    \\    
7 & 4.0$\pm$2.4        &3.2$\pm$0.6   &6797.51    &48.56(47.52,48.84) &35932.60   &O9/O9.5  &31628.14   &O9.5/B0  &35439.60   &O7.5/O8    \\    
9 & 1.3$\pm$0.6        &12.9$\pm$0.5  &14.85      &48.39(47.18,48.68) &34678.30   &O9.5/B0  &30751.53   &B0/B0.5  &33170.26   &O7.5/O8    \\    
13& 2.8$\pm$1.7        &3.7$\pm$0.6   &367.12     &48.88(47.86,49.16) &38595.87   &O8/O8.5  &33696.80   &O9/O9.5  &38924.94   &O7/O7.5    \\    
34& 1.2$\pm$0.6        &4.6$\pm$0.5   &474.90     &47.98(46.84,48.27) &32943.46   &B0/B0.5  & ------    & ------  &32608.77   &O8/O9      \\    
25& 3.6$\pm$2.9        &15$\pm$0.8    &219.12     &48.73(47.59,49.01) &37270.24   &O8.5/O9  &32643.71   &O9.5/B0  &37656.80   &O7/O7.5    \\    
27& 4.3$\pm$2.6        &14.2$\pm$0.6  &186.18     &48.77(47.63,49.05) &37570.05   &O8.5/O9  &32871.98   &O9.5/B0  &38185.60   &O7/O7.5    \\    
28& 3.9$\pm$1.6        &12.2$\pm$0.4  &123.24     &48.27(47.09,48.56) &33977.04   &O9.5/B0  &30160.00   &B0.5     &31603.38   &O7.5/O8    \\    
29& 2.2$\pm$1.1        &13.4$\pm$0.5  &283.27     &48.37(47.22,48.66) &34548.64   &O9.5/B0  &30641.69   &B0/B0.5  &32880.55   &O7.5/O8    \\    
30& 1.3$\pm$0.5        &4.5$\pm$0.4   &101.23     &47.28(46.07,47.57) & ------    & ------  & ------    & ------  &29569.05   &B0/B0.5    \\    
35& 1.9$\pm$0.8        &10.7$\pm$0.4  &201.72     &47.79(46.55,48.08) & ------    & ------  & ------    & ------  &31834.86   &O9.5/B0    \\    
38& 1.3$\pm$0.5        &10.4$\pm$0.4  &406.21     &48.84(47.74,49.12) &38191.86   &O7.5/O8  & ------    & ------  &39016.73   &O7.5/O8    \\    
48& 2.0$\pm$1.0        &9.5$\pm$0.5   &333.97     &48.24(47.08,48.53) &31657.87   &B0/B0.5  &29676.70   &O9.5/B0  &31241.88   &O7.5/O8    \\    
51& 3.1$\pm$1.9        &9.2$\pm$0.6   &236.20     &48.56(47.42,48.84) &35882.69   &O9/O9.5  &31585.72   &O9.5/B0  &35348.35   &O7.5/O8    \\    
54& 1.3$\pm$0.7        &9.7$\pm$0.5   &291.90     &47.66(46.40,47.95) & ------    & ------  & ------    & ------  &31271.76   &O9.5/B0    \\    
56& 0.5$\pm$0.2        &3.4$\pm$0.4   &708.97     &46.95(45.72,47.24) & ------    & ------  & ------    & ------  &27933.93   &B0.5/B1    \\    
57& 3.8$\pm$1.5        &10.3$\pm$0.4  &199.74     &48.54(47.41,48.82) &35740.28   &O9/O9.5  &31464.70   &O9.5/B0  &35088.02   &O7.5/O8    \\    
59& 4.6$\pm$2.3        &5.50$\pm$0.5  &184.85     &48.76(47.72,49.04) &37539.34   &O8/O8.5  &32828.97   &O9.5/B0  &35148.45   &O7.5/O8    \\    
62& 1.6$\pm$1.0        &8.9$\pm$0.6   &597.45     &48.62(47.49,48.90) &36351.62   &O8.5/O9  &31944.29   &O9.5/B0  &36036.56   &O7/O7.5    \\    
64& 1.4$\pm$1.3        &7.3$\pm$0.9   &363.17     &47.94(46.75,48.23) &33128.79   &B0/B0.5  & ------    & ------  &32464.46   &O8/O9      \\    
65& 1.0$\pm$0.9        &6.2$\pm$0.9   &786.50     &48.10(46.92,48.39) &32336.11   &B0/B0.5  & ------    & ------  &33081.67   &O8/O9      \\    
68& 5.4$\pm$4.3        &8.5$\pm$0.8   &95.57      &48.38(47.24,48.66) &34634.25   &O9.5/B0  &30714.21   &B0/B0.5  &34149.39   &O8/O9      \\    
69& 4.2$\pm$2.5        &13.2$\pm$0.6  &73.67      &48.00(46.73,48.29) &32834.86   &B0/B0.5  & ------    & ------  &32693.32   &O8/O9      \\    
71& 3.4$\pm$2.7        &10.8$\pm$0.8  &145.42     &48.28(47.10,48.57) &34021.96   &O9.5/B0  &30195.54   &B0/B0.5  &33745.98   &O8/O9      \\    
72& 4.7$\pm$2.4        &11.8$\pm$0.5  &175.81     &48.59(47.42,48.88) &36129.46   &O9/O9.5  &31795.45   &O9.5/B0  &35799.49   &O7.5/O8    \\    
73& 1.2$\pm$0.5        &10.6$\pm$0.4  &607.34     &48.09(46.89,48.38) &32416.96   &B0/B0.5  & ------    & ------  &33018.72   &O8/O9      \\    
74& 5.1$\pm$2.6        &10.1$\pm$0.5  &102.72     &48.52(47.40,48.80) &35615.44   &O9/O9.5  &31358.59   &O9.5/B0  &34859.78   &O7.5/O8    \\    
75& 2.7$\pm$1.3        &11.0$\pm$0.5  &119.97     &47.90(46.66,48.19) &33350.20   &B0/B0.5  & ------    & ------  &32292.06   &O8/O9      \\    
76& 2.7$\pm$1.4        &10.8$\pm$0.5  &266.81     &48.49(47.36,48.77) &35388.73   &O9/O9.5  &31234.94   &B0/B0.5  &34554.40   &O8/O9      \\    
78& 3.4$\pm$1.7        &10.1$\pm$0.5  &348.55     &48.68(47.62,48.96) &36865.30   &O8.5/O9  &32335.39   &O9.5/B0  &35296.18   &O8/O9      \\    
79& 1.6$\pm$0.9        &11.5$\pm$0.6  &676.56     &48.38(47.20,48.67) &34623.36   &O9.5/B0  &30704.99   &B0/B0.5  &34142.21   &O8/O9      \\    
80& 6.2$\pm$3.1        &9.2$\pm$0.5   &59.35      &48.20(47.03,48.49) &33574.19   &O9.5/B0  & ------    & ------  &33450.96   &O8/O9      \\    
82& 0.4$\pm$0.2        &4.2$\pm$0.6   &895.04     &47.13(45.89,47.42) & ------    & ------  & ------    & ------  &28778.41   &B0/B0.5    \\    
83& 1.0$\pm$0.7        &1.3$\pm$0.7   &263.36     &47.13(46.11,47.41) & ------    & ------  & ------    & ------  &28798.63   &B0/B0.5    \\    
85& 1.4$\pm$7.2        &1.3$\pm$0.6   &79.53      &48.75(47.61,49.04) &37442.09   &O8/O8.5  &32760.09   &O9.5/B0  &35049.68   &O7.5/O8    \\    
86& 4.4$\pm$0.4        &6.1$\pm$0.1   &52.19      &47.65(46.49,47.94) & ------    & ------  &  ------   & ------  &31243.40   &O9.5/B0    \\    
87& 5.9$\pm$0.6        &6.1$\pm$0.1   &53.62      &48.27(47.17,48.55) &33968.53   &O9.5/B0  &30160.00   &B0.5     &33710.77   &O8/O9      \\    
88& 1.0$\pm$2.2        &6.0$\pm$2.2   &537.61     &47.69(46.54,47.98) & ------    & ------  & ------    & ------  &31398.63   &O9.5/B0    \\    
89& 3.9$\pm$4.3        &16.2$\pm$1.1  &157.46     &48.62(47.44,48.91) &36381.57   &O8.5/O9  &31967.09   &O9.5/B0  &35062.57   &O8/O9      \\    
90& 5.5$\pm$5.5        &5.8$\pm$1.0   &52.62      &48.05(46.93,48.33) &32580.67   &B0/B0.5  & ------    & ------  &32891.25   &O8/O9      \\    
91& 3.5$\pm$2.4        &10.2$\pm$0.7  &146.99     &48.36(47.20,48.65) &34484.99   &O9.5/B0  &30587.77   &B0/B0.5  &34051.05   &O8/O9      \\    
\hline
\end{longtable}

\newpage
\section{Spectral-type estimates for the ionizing stars}

The results of determining the 20~cm radio continuum fluxes toward 91 objects from the catalog~\cite{Topchieva} are presented in Table~\ref{tab:catal8mkm}. To determine the spectral types of the corresponding stars, we first estimated the UV flux ionizing the hydrogen ($Q_{\rm Ly}$), assuming that the radio emission is optically thin. This is valid for ``classical'' HII regions (e.g.,~\cite{Dirienzo, Condon}). We used the following expression from~\cite{Dirienzo} (see also~\cite{Condon}, Tsivilev et al.~\cite{Tsivilev}):

\numberwithin{equation}{section}
\begin{equation}
Q_{\rm Ly}\gtrsim7.54\times10^{46}\biggl(\frac{T_{\rm e}}{10^{4} \rm{K}}\biggl)^{-0.45}\biggl(\frac{\nu}{\rm GHz}\biggl)^{0.1}\biggl(\frac{F_{\rm 20}}{\rm Jy}\biggl)\biggl(\frac{D}{\rm kpc}\biggl)^{2} s^{-1},
\end{equation}

where $T_{\rm e}$ is the electron temperature, taken to be $\thickapprox 10^4$, as is typical for HII regions (e.g.,~\cite{Tielens}), $\nu = 1.49$~GHz is the frequency corresponding to 20~cm wavelength, $F_{\rm 20cm}$ the flux at 20~cm, and $D$ the distance to the HII region. The value of $Q_{\rm Ly}$ determined in this way represents a lower limit to the flux of ionizing UV photons, since some UV photons could penetrate outward through rarified parts of the molecular clouds due to inhomogeneity of the matter in the ISM. We did not take this into account when carrying out our estimates (see~\cite{Anderson_0}).

We estimated the spectral types from the available values of $Q_{\rm Ly}$ using the results of theoretical computations by Vacca et al.~\cite{Vacca} and Smith et al.~\cite{Smith}. The results of~\cite{Vacca}] enable estimation of $Q_{\rm Ly}$ for all 42 objects in our sample for which distance estimates are available, assuming that these regions are formed only by stars having luminosity class V. This is not always valid, since young stars ionizing HII regions may be giants~\cite{Avedisova, Alexeeva, Walborn, Walborn_1}. Therefore, we used the tables of~\cite{Vacca} for luminosity classes III and V. We did not consider luminosity class Ia, since such stars are rarely encountered among stars ionizing HII regions, and our estimates of $Q_{\rm Ly}$ do not fall in the range presented for this luminosity class in~\cite{Vacca}. We determined the electron densities in the HII regions based on the objects' distances and angular sizes (see~\cite{Topchieva}) using the following expression from~\cite{Panagia}:

\numberwithin{equation}{section}
\begin{equation}
n_{\rm e} = 3.1113\times10^{2} C_{\rm 1}\biggl(\frac{F_{\rm 20}}{\rm Jy}\biggl)^{0.5}\biggl(\frac{T_{\rm e}}{10^{4} {\rm K}}\biggl)^{0.25}\biggl(\frac{\rm D}{\rm kpc}\biggl)^{-0.5} b({\nu},T)^{-0.5} {\theta}_{\rm R}^{-1.5} cm^{-3},
\end{equation}

where ${\theta}_{\rm R}$ is the angular size in arcminutes, С$_{\rm 1}$ is a constant, equal to 0.8165 for a uniform sphere~\cite{Panagia}, and the coefficient $b({\nu},T)$ was calculated using the formula

\numberwithin{equation}{section}
\begin{equation}
b({\nu},T) = 1+0.3195 {\rm log_{10}}\biggl(\frac{T_{\rm e}}{10^{4} {\rm K}}\biggl)-0.2130 {\rm log_{10}}\biggl(\frac{\nu}{1 {\rm GHz}}\biggl),
\end{equation}

The spectral types of the ionizing stars and parameters of the HII regions are presented in Table~\ref{tab:mainres}. The spectral types range from B0.5 to O7, which corresponds to effective temperatures from 29 000 to 37 000 K.

\begin{figure}[h!]
\includegraphics[width=0.8\linewidth]{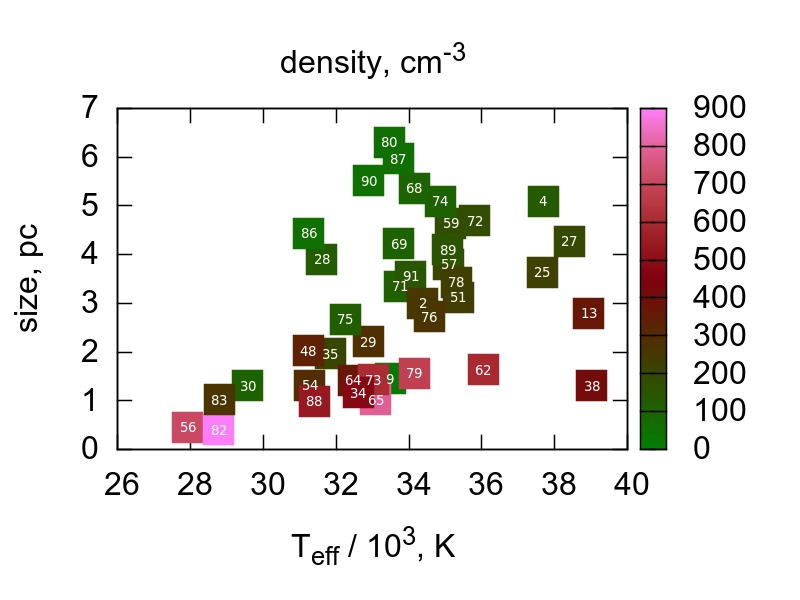}
\caption{Proposed evolutionary chains for HII regions. The horizontal axis plots $T_{\rm eff}$ and the vertical axis the size of the HII regions, with the colors corresponding to the electron densities in the HII regions. The numbers identify the objects according to the numbers in Tables 1 and 2. The symbols for objects 2, 9, 65, 73, and 80 have been shifted slightly along the horizontal axis to make their numbers visible.}
\label{fig:bubbleevolve}
\end{figure}

Figure~\ref{fig:bubbleevolve} shows the dependence of the electron density ($n_{\rm e}$) on the size of the HII regions and the effective temperatures $T_{\rm eff}$ of the ionizing stars using a color scale. This shows clearly that, when $T_{\rm eff}$ is fixed, $n_{\rm e}$ is lower for larger HII regions. An exception is the region N4, which has a complex morphology, and we do not consider this region further here.

The results shown in Fig.~\ref{fig:bubbleevolve} correspond to the idea that the size of an HII region increases in the course of its evolution, while its electron density decreases. This is supported by general theoretical considerations and numerical modeling~\cite{Shu, Kirsanova_09}. Thus, we can distinguish chains of HII regions in our catalog located at different stages in their evolution.

\section{Comparison of the radio and IR fluxes}\label{sec:ir}

In this section, we analyze the relationship between the radio and IR fluxes. The radiation at 8 $\mu$m corresponds to emission by PAHs, at 24 $\mu$m to emission by small, hot dust grains, and at 160 $\mu$m to emission by large, fairly cool dust grains~\cite{Draine, 2007ApJ...657..810D}.

\begin{figure}[h!]
\includegraphics[width=0.5\linewidth]{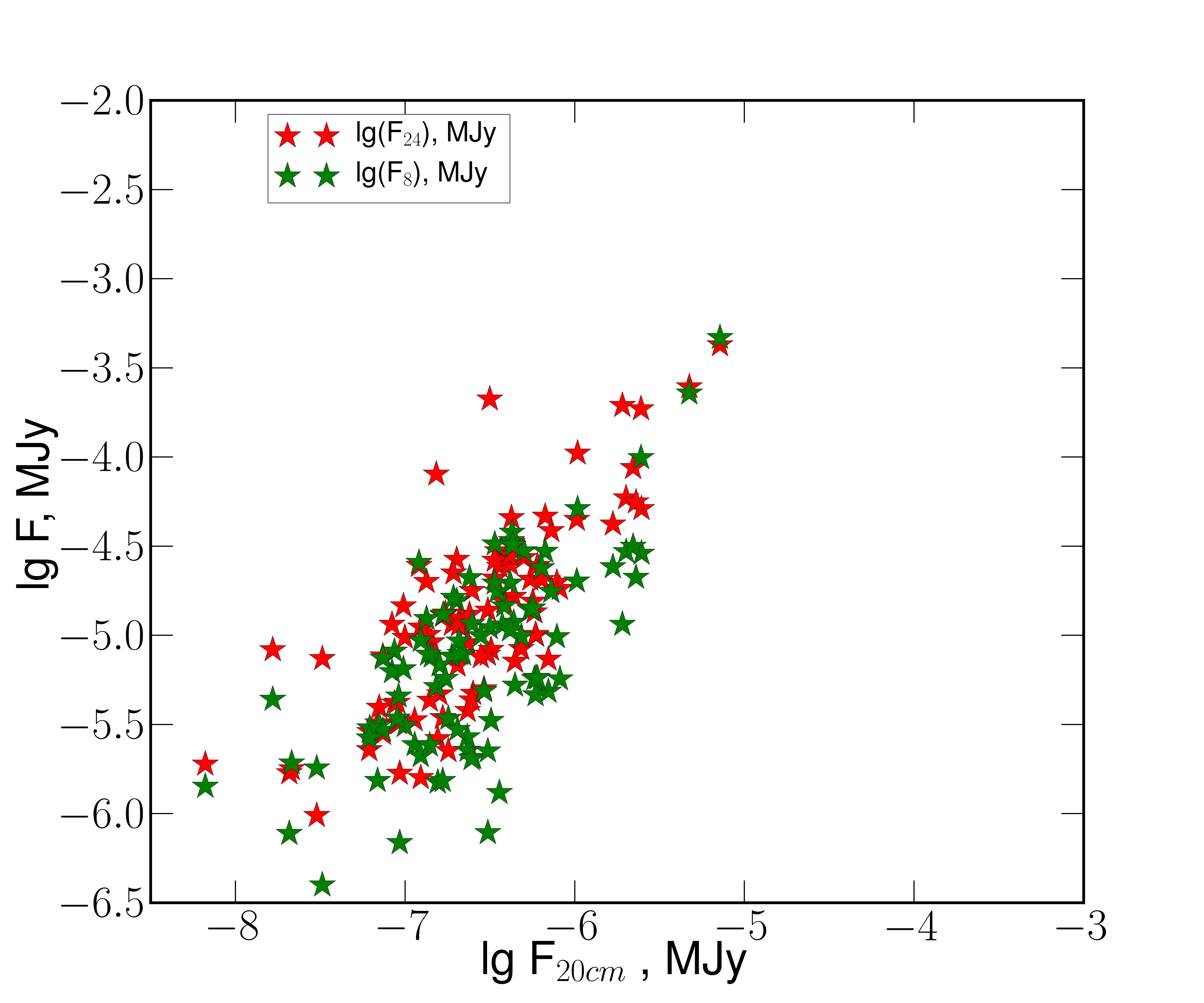}

\caption{Dependence of the IR flux on the radio flux for 91 objects. Green symbols show data for 8 $\mu$m and red symbols data for 24 $\mu$m.}
\label{fig:1_2}
\end{figure}

Figure~\ref{fig:1_2} shows the dependence of the 8 and 24 $\mu$m IR fluxes on the 20~cm radio flux. The data presented in Fig.~\ref{fig:1_2} show that higher radio continuum fluxes correspond to higher IR fluxes.

\begin{figure}[h!]
\includegraphics[width=0.35\linewidth]{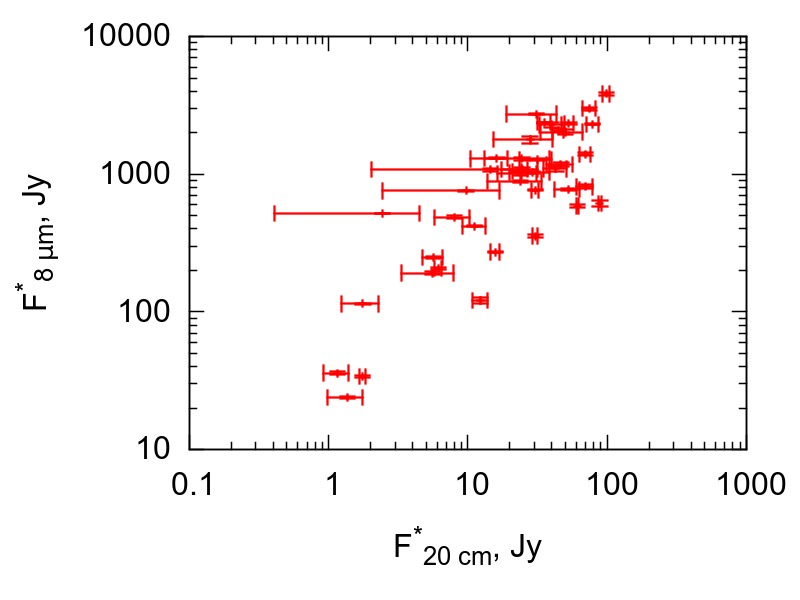}
\includegraphics[width=0.35\linewidth]{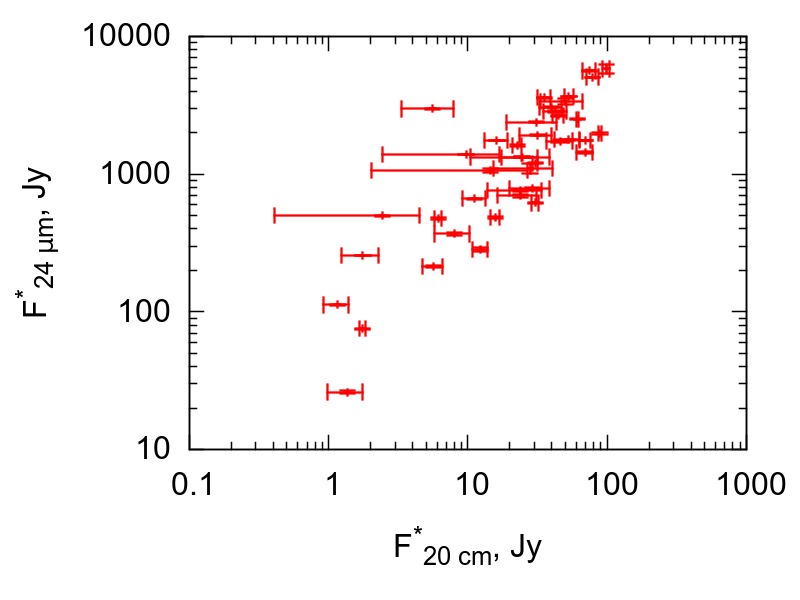}
\includegraphics[width=0.35\linewidth]{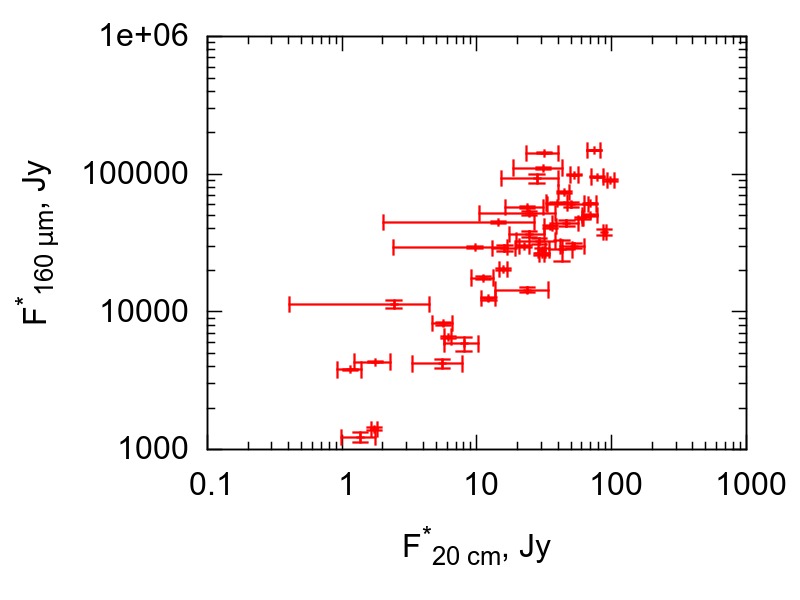}
\caption{Dependence of the IR flux on the radio flux after taking into account the distances to the objects via the normalizing coefficients $1/D^{2}$.}
\label{fig:fluxerror}
\end{figure}

To exclude the influence of differences in the distances, we multiplied the fluxes by $1/D^{2}$. The resulting dependences of the 8, 24, and 160 $\mu$m IR fluxes on the 20~cm radio flux is shown in Fig.~\ref{fig:fluxerror}.

The scatter in these dependences is due mainly to uncertainties in $F_{\rm 20cm}$. The uncertainties in the fluxes were calculated from the dispersions of the fluxes for a background area of sky in the same frame (the choice of aperture for calculating this background is described in~\cite{Topchieva}). The contribution of instrumental uncertainties does not exceed 0.5\%, and we did not take this into account. The uncertainties in the IR fluxes are much lower, and cannot be distinguished in Fig.~\ref{fig:fluxerror}.

We can see that the IR fluxes grow roughly linearly with the radio flux. The dependence of $F_{\rm 8 {\mu}m}$ on $F_{\rm 20cm}$ has the largest scatter. These plots show that the higher the radio flux (the earlier the spectral type of the star), the higher the IR flux.

The IR flux is determined by two quantities --- the mass and temperature of the dust envelope around the HII region~\cite{Mathis}. The 8 and 24 $\mu$m emission is formed in the falling part of the spectrum of ``warm'' dust. In this case, it is not possible to distinguish whether it is variations in the mass or temperature that are responsible for the dependence of $F_{\rm 8 \mu m}$ and $F_{\rm 24 \mu m}$ on $F_{\rm 20cm}$.

The emission at 160 $\mu$m is associated with ``cool'' dust, which forms a well defined peak around 100-200 $\mu$m in the spectra of many HII regions. In this case, the 160 $\mu$m emission is formed almost exclusively in cool dust with temperatures of about 18-22 K~\cite{Draine, 2007ApJ...657..810D}. The fraction of such regions in our sample is about 70\%~\cite{Topchieva}. Thus, the increase in the 160 $\mu$m flux with growth in the effective temperature of the ionizing star may mean that hotter stars are surrounded by cool dust envelopes with higher masses.

\begin{figure}[h!]
\includegraphics[width=0.35\linewidth]{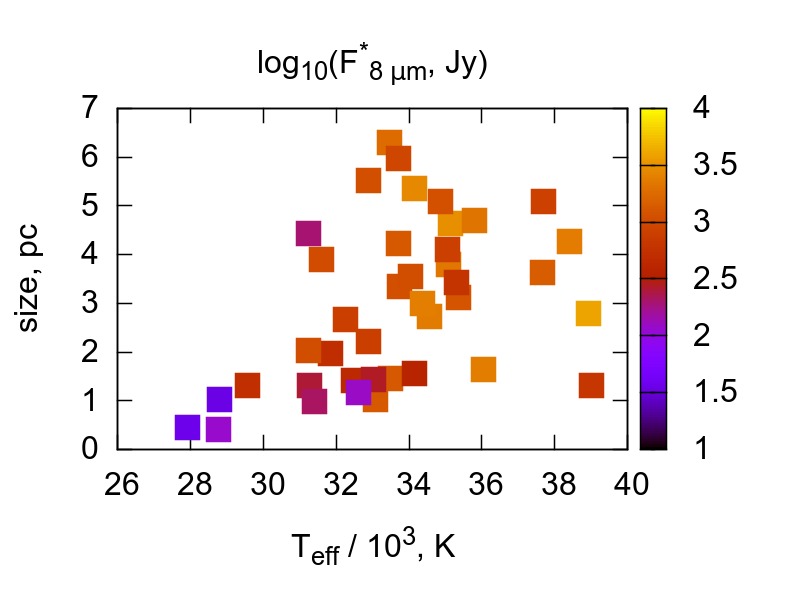}
\includegraphics[width=0.35\linewidth]{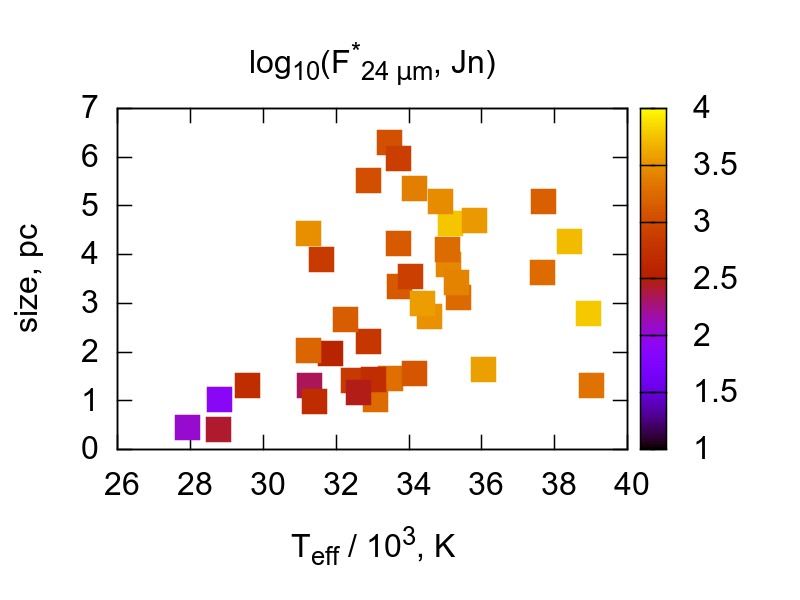}
\includegraphics[width=0.35\linewidth]{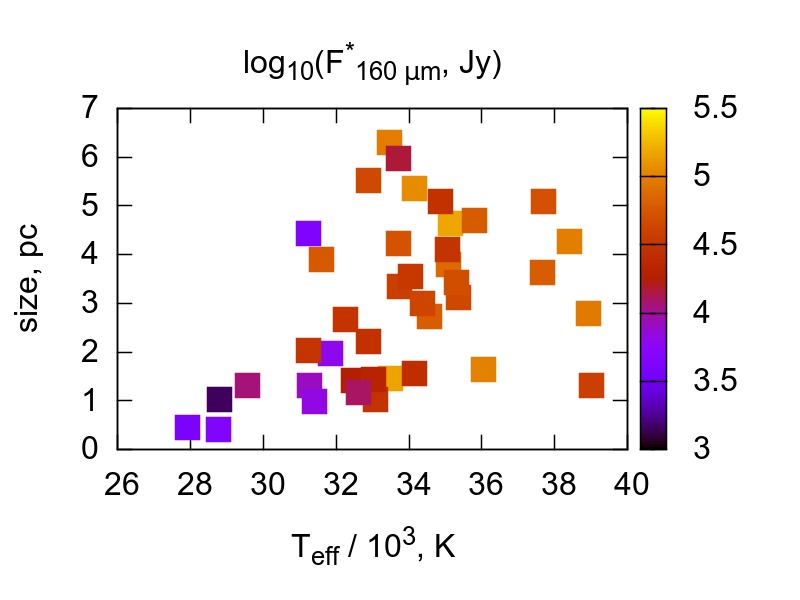}
\caption{Plots of the 8, 24, and 160 $\mu$m fluxes as functions of the sizes of the HII regions and effective temperature of the ionizing stars, with the flux densities shown on the indicated color scale.}
\label{fig:fluxevolve}
\end{figure}

Figure~\ref{fig:fluxevolve} shows the dependences of the IR fluxes (shown by the color scale) on the effective temperatures of the ionizing stars and the sizes of the HII regions. There is no dependence of the fluxes on the sizes of the regions when a fixed effective temperature is considered. Analysis of the results presented Fig.~\ref{fig:bubbleevolve} shows that the size of an HII region for fixed $T_{\rm eff}$ is an indicator of age. Thus, no clear dependence (growth or decrease) of the IR fluxes on age is traced in our data. This may mean that the dynamics of the expansion of the HII regions differs appreciably from the predictions of idealized theoretical models (e.g.,~\cite{Spitzer}); for example, due to inhomogeneity in the initial distribution of the gas in the molecular clouds or the development of inhomogeneities due to thermal instability (e.g.,~\cite{Parker, Osterbrock}; see also~\cite{Krasnobaev} concerning the development of thermal instability in PDRs). This may explain why there is no continuous growth in the masses of envelopes around HII regions with time, as is suggested, for example, by Eqs. (39)-(43) of the analytical computations of Hosokawa et al.~\cite{Hosok}. The penetration of UV photons outward through the HII region, for which the efficiency is shown, for example, by Ferguson et al.~\cite{Ferguson} and Oey et al.~\cite{Oey}, apparently occurs through the least dense regions in the molecular clouds. This process is accompanied by a reduction in the transformation of the radiative energy of the massive stars into the kinetic energy of moving gas, a slowing of the growth of the mass of the envelope around the HII region, and a decrease in heating of the envelope with time. Since the HII regions for which data are presented here are not observed in optical lines such as H$_{\alpha}$, we suggest that these objects are deeply embedded in their host molecular clouds. However, diffuse IR emission at 8 $\mu$m is present around these objects, since the UV photons responsible for the PAH flourescence probably penetrate through gaps in dense parts of the envelope. This penetration of ionizing photons is probably negligible in the HII regions in our sample, since these objects are not visible in the optical wavelengths, indicating they are embedded in the molecular gas. Therefore, we conclude that it is apparently the inhomogeneous distribution of the gas that leads to the absence of a dependence of the IR fluxes on the ages of the HII regions.

\section{Conclusion}\label{sec:results}

We have obtained the following results in this study:

\begin{itemize}
\item We have determined the 20~cm radio continuum fluxes toward 91 HII regions.
\item We have determined the spectral types for the ionizing stars in 42 objects with available distance estimates. These spectral types range from B0.5 to O7, corresponding to effective temperatures from 29 000 to 37 000 K.
\item We have established that the IR fluxes at 8, 24, and 160 $\mu$m grow with the 20~cm flux. The 160 $\mu$m flux increases with the effective temperature of the ionizing star. This may indicate that hotter stars are surrounded by cool dust envelopes with higher masses. In the case of the 8 and 24 $\mu$m emission, it is not possible to distinguish the influences of the masses and temperatures of the dust envelopes around the HII regions on the IR fluxes $F_{\rm 8 \mu m}$ and $F_{\rm 24 \mu m}$.
\item There is no dependence of the IR fluxes on the sizes of the HII regions when the effective temperature of the ionizing star is fixed. Our results may indicate an absence of a continuous growth in the masses of the envelopes around the HII regions with time due to nonuniformity in the initial distribution of gas in the ISM and/or the escape of ionizing UV photons through gaps in the envelopes.
\end{itemize}

We thank V.V. Akimkin, D.Z. Wiebe, and the anonymous referee for useful comments that have helped us improve the quality of this paper. This work was supported by the Russian Foundation for Basic Research (grant nos. 17-32-50058 and 18-02-00917). This work has made use of the Astropy software package~\cite{Astropy}.

\selectlanguage{english}

\end{document}